\documentclass[prl,aps,twocolumn, superscriptaddress,showpacs]{revtex4}
\usepackage{graphicx, amsmath, amsthm}
\makeatletter
\begin{document}
\title{Impurity effects on the Quasi-particle spectrum in the
Fulde-Ferrell-Larkin-Ovchinnikov state of a $d$-wave superconductor}

\author{Qian Wang}
\affiliation{Texas Center for Superconductivity, University of Houston,
Houston, Texas 77204}
\author{Chia-Ren Hu}
\affiliation{Department of Physics, Texas A\&M University, College
Station,
Texas 77843}
\author{Chin-Sen Ting}
\affiliation{Texas Center for Superconductivity, University of Houston,
Houston, Texas 77204}
\date{\today}
\begin{abstract}
In a $d$-wave superconductor (d-SC), a unitary impurity can induce a
near-zero-energy resonant peak in the local tunneling density of
states (LTDOS) due to the sign change of the order parameter (OP)
on the Fermi Surface.  If a d-SC is quasi-two-dimensional, a large
parallel magnetic field can drive it into the
Fulde-Ferrell-Larkin-Ovchinnikov (FFLO) state, with an OP also
changing sign in the real space like a checkerboard pattern.
This ``double sign change'' leads to very subtle effects by a
unitary impurity on the LTDOS for two locally stable locations
of the impurity.
\end{abstract}
\maketitle

In a homogeneous, non-$s$-wave, singlet superconductor (SC), a unitary,
non-magnetic impurity can induce several quasi-localized, near-zero-energy
(relative to the Fermi energy) resonant states (NZERSs), which are the
direct consequence of the sign change of the superconducting (SCing) order
parameter (OP) on the Fermi surface~\cite{Balatsky,Balatsky95,%
Pan00,Zhu00,Zhu-Ting-Hu,MartinBalatskyZaanen}. These states
are responsible for the pair-breaking effects of non-magnetic
impurities in unconventional (i.e., non-$s$-wave) SCs, and they can
lead to a near-zero-bias resonant peak (NZBRP) in the local
tunneling density of states (LTDOS) near the impurity.
This peak appears near the minimum of the bulk DOS of such a SC,
and is one of the clearest evidence for unconventional pairing
in, for example, high-$T_c$ SCs.  These resonant states are close
kin of the so-called zero-energy Andreev bound states (ZEABS, also
known as the midgap states~\cite{hu94}) which form at properly-oriented
surfaces and interfaces of non-$s$-wave SCs, and are responsible for
the zero-bias conductance peaks (ZBCPs) observed ubiquitously in
various types of tunneling experiments performed on various kinds
of unconventional SCs. These ZEABSs are also the direct consequence
of the sign change of the OP on the Fermi surface, and have in fact
a topological origin~\cite{topology}. On the other hand, If a SC is
quasi-two-dimensional, a strong magnetic field applied parallel to
its layers can cause a large Zeeman splitting between its spin-up
and -down electrons, and the orbital effect of the magnetic field
can be suppressed. Then an inhomogeneous SCing state known as the
Fulde-Ferrell-Larkin-Ovchinnikov (FFLO) state~\cite{FFLO} can become
energetically more favorable. We have recently shown~\cite{qwang05}
that in an $s$-wave SC (s-SC), the OP of the FFLO state changes sign
periodically in real space along one direction, with a periodic array
of parallel real-space nodal lines, whereas in a $d_{x^2-y^2}$-wave
SC (d-SC), the FFLO state changes sign periodically in two mutually
perpendicular directions, forming a checkerboard pattern, with two
mutually perpendicular sets of parallel real-space nodal lines,
that are along the nodal directions the d-wave OP in the (relative)
momentum space, i.e., at 45$^\circ$ with the $a$- and $b$-axes. 
Right along these real-space nodal lines, but away from the saddle
points where real-space nodel lines cross, ZEABSs can also form for
the same topological reason~\cite{qwang05}.  Then, as isolated
impurities are added into a d-SC in such a state, two types of
near-zero-energy quasi-particle states can potentially form --- the
ZEABSs localized near the real-space nodal lines, and the NZERSs
localized near the impurities~\cite{note1}. The possible mutual
interaction of these two types of states then constitute an
extremely interesting and fundamental topic, if only that these
two types of states are not far apart, so that their wave-functions
can overlap. Questions that can be raised include: Do they produce
overlaping peaks in the LTDOS? Or there is some sort of level
repulsion, but then how? (Would both types move to finite energies,
or just one type?) Can one type preclude the existence of the other
type? We shall see that the answer depends on the location of the
impurity, and the results are quite unexpected. Confirming these
results experimentally should then constitute one of the strongest
evidences for the FFLO state in a d-SC. The FFLO state may
very-likely have been realized in the heavy fermion compound
CeCoIn$_5$~\cite{FFLO-CeCoIn5}, which is very likely a
d-SC~\cite{dSC-CeCoIn5}, although no evidences presented so far
for the FFLO state are direct ones.

To perform such a theoretical study, we need to first determine the
locally stable locations of the impurity. (At very low concentrations
of impurities, the spatial structure of the OP will destort slightly
so that all impurities will be at such locations. It will not be
true at higher concentrations, which will be studied in a future work.)
We again solve the discrete Bogoliubov-de Gennes equations as
before~\cite{qwang05}:
\begin{equation}
\sum_j\left({{\cal H}_{ij,\sigma}\atop \Delta_{ij}^*}
        {\Delta_{ij}\atop{-{\cal H}_{ij,\bar\sigma}^*}}\right)
        \left(u_{j\sigma}^n\atop v_{j\bar\sigma}^n\right) =
        E_n\left(u_{j\sigma}^n\atop v_{j\bar\sigma}^n \right),
\label{discreteBdG}
\end{equation}
except that here
$H_{ij,\sigma} = -t - (\mu+\sigma h)\delta_{ij}+U_0\delta_{i,j_0}$
contains an impurity of strength $U_0$ located at $j_0$.
$u_{j\sigma}^n$ and $v_{j\bar\sigma}^{n}$ are the Bogoliubov
quasiparticle amplitudes on the $j$-th site. The self-consistency
condition for the OP:
\begin{equation}
\Delta_{ij} =
\delta_{j,i+\gamma}\frac{V}{4}\sum_n\tanh\frac{E_n}{2k_BT}(u^n_{i\uparrow}
    v^{n*}_{j\downarrow}+u^n_{j\downarrow}v^{n*}_{i\uparrow})\,.
\label{selfconsist}
\end{equation}
is solved by iteration. Here $\gamma = (\pm 1,0)$ and $(0,\pm 1)$, and
$\Delta_i=(\Delta_{i+\hat x} +
\Delta_{i-\hat x} - \Delta_{i+\hat y} - \Delta_{i-\hat y})/4$
is the d-SC OP at site $i$.

Previously we have studied the FFLO state in a clean system~\cite{qwang05}.
We use one of the solutions as the initial configuration (but with
all $\Delta_{j_0,j_0+\gamma}$ set to 0). Both strong unitary (i.e.,
$U_0 = 100$) and weak ($U_0 = 1$) impurities have been investigated.
We set $V=1.0$, $\mu=-0.4$ and $h=0.15$.  The OP in a clean system has
been plotted as Fig. 1(b) in Ref.~\cite{qwang05}.

In the presence of an impurity the OP still iterates practically
to the same 2D lattice in a d-SC, except for a local depression
at the impurity site.
% as shown in Fig.~\ref{fig:order}.
The size of the order parameter
hole created is of the order of the coherence length.
Depending on the initial configurations, the iteration leads either
to an OP saddle point being pinned at the impurity
site,
%as shown in Fig.~\ref{fig:order}(a),
or, with slightly higher energy, an OP extremum, located at the center of a basic OP lattice bounded by 4 nodal lines, being pinned at the
impurity site~\cite{note2}.
%as shown in Fig.~\ref{fig:order}(b)
When at an OP saddle point, the OP essentially vanishes at more sites
around the impurity than when there is no impurity. When at an OP
extremum, the OP vanishes on that site and is suppressed at its
surrounding sites. (At the impurity site the OP practically vanishes
for $U_0=100$, and is only suppressed by about 30\% for $U_0=1$).

% \begin{figure}
% \centering
% \includegraphics[width=2.5in] {order.eps}
% \caption{ Spatial variations of $\Delta_i$ for a unitary impurity
% (located by a red arrow) in a d-SC in the FFLO state.
% In (a), the impurity is located at an OP saddle point;
% In (b), the impurity is located at an OP extremum of the pure system.}
% \label{fig:order}
% \end{figure}

Next, we calculate the LDOS near the impurity.
The LDOS of spin-up and -down quasi-particles is given by:
\begin{equation}
\rho_{i\sigma}(E)=\sum_n [|u^n_{i\sigma}|^2\delta(E_n-E) +
|v^n_{i\bar\sigma}|^2\delta(E_n+E)]\,.
\end{equation}
In what follows, we only present the results for spin-up
quasi-particles since the spin-down LDOS spectra are simply
the spin-up ones shifted to the right by 2$h$. The sum of
spin-up and -down LDOS gives the total local differential
tunnel conductance, measured by Scanning Tunneling Microscopy
(STM) with an unpolarized tip.

\begin{figure}[htb]
\centering
\includegraphics[width=2.5in]{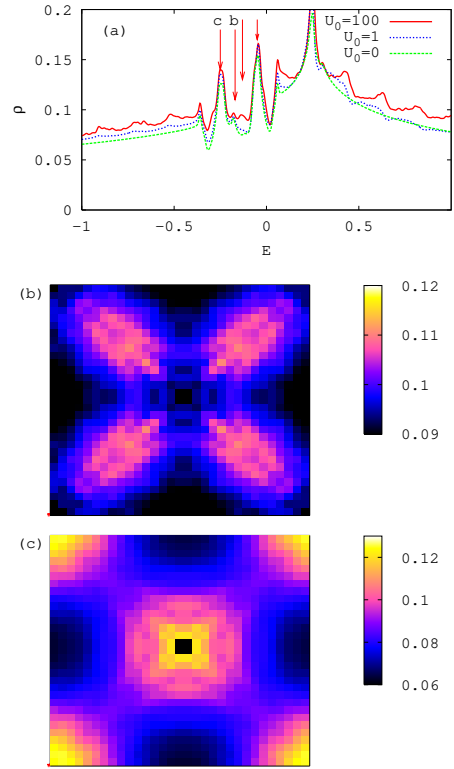}
\caption{(a): The spin-up LDOS spectrum on a nearest-neighbor
site of the impurity, which is at an OP saddle point.
The subgap peaks of the LDOS are marked by arrows. (b) and (c):
The LDOS maps at the peak-energies $E=-0.17$ and $E=-0.25$,
respectively, for $U_0 = 100$. In both maps,
the impurity is at the center. The four corners of
these maps are the neighboring saddle points of the OP.}
\label{fig:ldos}
\end{figure}

When an OP saddle point is pinned at the impurity site,
Fig.~\ref{fig:ldos} (a) shows the spin-up LDOS on a nearest-neighbor
site of the impurity, revealing the same four subgap peaks (marked by
arrows, and confirmed by the maps in parts (b) and (c) of this figure),
as discussed in Ref.~\cite{qwang05}, as well as the coherence peaks
and a van Hove peak, but no new peak(s) that can be identified as the
impurity-induced resonant peak(s)~\cite{Balatsky}.  Comparing
this figure with the LDOS plot at an OP saddle point in Fig.~4(b) of
Ref.~\cite{qwang05}, we see that the outer two subgap peaks are only
perturbed very weakly by the presence of a unitary impurity.
(High-energy oscillations are caused by quasi-particles trapped between
the impurities in the neighboring super-cells. They weaken as $U_0$ is
reduced.)

At first sight these results appear puzzling: According to
Ref.~\cite{qwang05}, the inner two weak subgap peaks are due to
the tails of the wave functions of ZEABSs, or midgap
states~\cite{hu94}, localized in this case near the halfway
points between the center OP saddle point and its
neighboring OP saddle points. Thus their weak dependence on
the impurity potential near the center saddle point is not surprising.
But the outer two strong subgap peaks can be identified as due to
the finite-energy ABSs localized at the center saddle
point~\cite{qwang05}. Why do these peaks also depend very weakly
on the impurity potential right at this saddle point? Also, why
are there no new resonant peaks induced by this impurity? Answer
to these questions lies in the difference between the ABSs due to
an OP well and the usual bound states by an ordinary potential
well. For the latter, adding a strong impurity potential
at the center of the potential well will certainly shift
the energy of the bound state, but for the former, the
quasi-particle is essentially moving at Fermi momentum,
and can be described by a semi-classical orbit. Without the impurity
the orbit is a straight line segment shooting through the OP saddle
point, with both ends terminated by Andreev reflections
involving the same sign of the OP. The actual bound state is
a coherent superposition of all such classical orbits of different
orientations. With a unitary impurity at the saddle point, the
classical orbit is deflected by the impurity to a new arbitrary
direction (in coherent superposition) but the new OP-value encountered
has either two plus signs or two minus signs (one in momentum space
since the OP is $d$-wave, and one in real space due to the FFLO state),
leading to always no net sign change of the pair potential. After
coherent superposition of all initial directions of the classical
orbit, one can see that the resultant semi-classical bound-state
wave function must be practically unchanged. (For a weaker impurity
potential, the deflection probability is reduced, but the conclusion
about the wave function remains practically the same.) This explains
the weak dependence of the outer two subgap peaks on $U_0$, and why
there are no impurity-induced NZERSs in this case, which requires
seeing a sign change of the OP.

\begin{figure}[htb]
\centering
\includegraphics[width=2.5in]{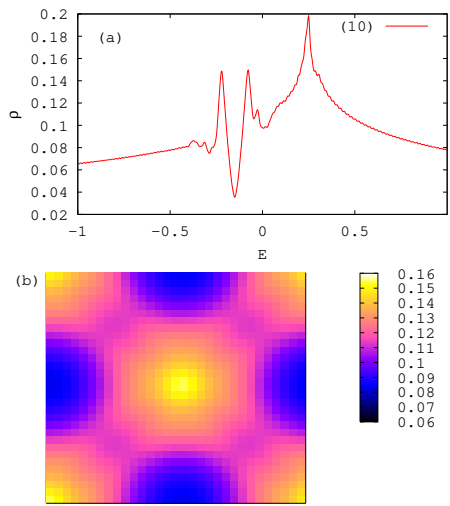}
\caption{(a): Similar to Fig.~\ref{fig:ldos} except that the real-space
sign change of the OP has been removed, and there is no impurity. 
In (a), the LDOS is calculated at a saddle point. Only
two subgap peaks now appear near the two outer subgap peaks obtained
in Ref.~\cite{qwang05}. (b): The LDOS map at one of the subgap peaks
($E=-0.22$).}
\label{fig:AbsDelta-ldos}
\end{figure}

In order to test this reasoning, we change the OP to its absolute
value (i.e., removing the real-space sign change but keeping the
characteristics of the d-SC), and re-calculated the LDOS at sites
next to the original OP saddle point and its map, as shown in
Fig.~\ref{fig:AbsDelta-ldos}.
Here only the two outer subgap peaks found in Fig.~\ref{fig:ldos} (a)
appear, as confirmed by the map in Fig.~\ref{fig:AbsDelta-ldos}(b) which shows
the maximum intensity at a saddle point,
but with slightly shifted energies, showing that these quasi-particle
states do not result from the sign change of the OP in real space.
Adding a unitary impurity to the center saddle point (not shown)
actually makes the ``impurity-induced near-zero-energy resonant
peak reappear, because the semiclassical scattering orbit can now
encounter both signs of the OP.

\begin{figure}[ht]
\centering
\includegraphics[width=2.5in]{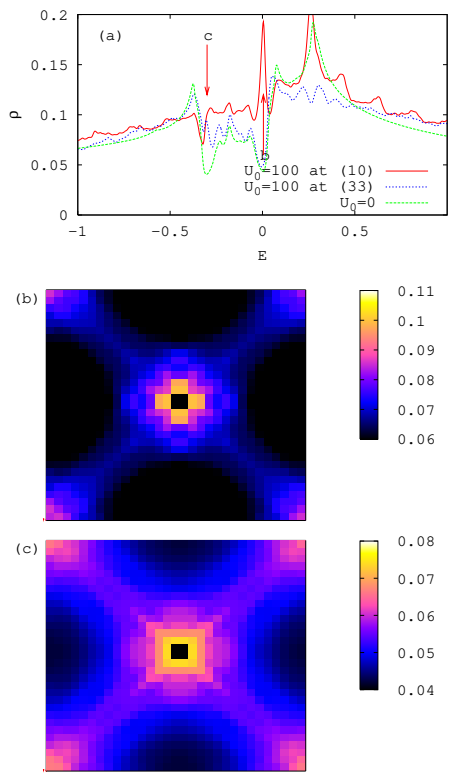}
\caption{(a): The spin-up LDOS spectrum at the (10) and (33) sites
relative to a unitary impurity at the (00) site, which is here also
an OP extremum, revealing two impurity-induced resonant peaks,
indicated by arrows marked for maps in (b) and (c), which confirm
this interpretation. The $U_0 = 0$ curve is for a pure system. (b)
and (c): The LDOS map at peak-energies $E=0.005$ and $E=-0.305$,
respectively, for $U_0=100$. In these maps, the impurity is located
at the center. The four corners are maximum $|\Delta|$ sites in the
neighboring cells of the 2D FFLO state.}
\label{fig:ldos1}
\end{figure}

Next we consider when an impurity is located at an OP extremum.
In this case we might naively think that a unitary impurity can
induce NZERSs in the LDOS just as in a uniform d-SC~\cite{Balatsky},
since the impurity is in a local environment where the OP has a
single sign in the real space, and only changes sign in the (relative)
momentum space. However, Fig.~\ref{fig:ldos1}(a) shows that two
resonant peaks are induced by the unitary impurity at the LDOS minima
or energy quasi-gaps of the host, and they are a pair separated by
$\pm\epsilon_0$  before the Zeeman shift, with $\epsilon_0$ about
68\% of the maximum gap for the parameter values considered.
A strong resonance peak due to the impurity located on the right of
the subgap states of the pure FFLO state could be clearly seen at
the (11) site relative the impurity at the (00) site. The left resonant
peak induced by the impurity is better revealed at the (33) site,
where the other subgap peaks are lower. These resonant states are
essentially localized around the impurity as is shown by the spatial
maps calculated at these peak biases shown in Fig.~\ref{fig:ldos1}
(b) and (c). For a weaker impurity these impurity-induced peaks are
even closer to the coherence peaks. These peaks are clearly located
outside the subgap peaks of a pure FFLO state~\cite{qwang05}. Those
subgap peaks are low in this plot as expected, since they are strong
near the OP saddle points and halfway points between the neighboring
saddle points only.

Separate calculations (with details omitted here) show that the
unitary-impurity-induced resonant peaks are also at finite energies
for $\Delta({\bf r}) = \Delta_0\,\cos ({\bf q}\cdot{\bf r})$,
but are very near zero energy for
$\Delta({\bf r}) = \Delta_0\,\exp (i{\bf q}\cdot{\bf
r})$~\cite{zhu-ting-chu}.
We offer the explanation below: The impurity-induced states are
resonant states because their energies are outside the gap on the
part of the Fermi surface near the nodal directions in the momentum
space. But relative to the rest of the Fermi surface they are simply
bound states localized near the vicinity of the impurity. It is
well-known that bound states can only form in the forbidden gaps of the
continuum states. Now for the FFLO states, whether 1D or 2D, there is
already a band of continuum states centered at the gap center\cite{qwang05}.
This band is narrower for field closer to the lower critical field
of the FFLO state, allowing the impurity-induced resonant peaks to
also move toward the gap center. Also, this band is clearly absent
in the current-carrying FF state, allowing the impurity-induced
resonant peak to still appear as a NZBRP before the Zeeman shift.

Fig.~\ref{fig:ldos1}(b) and (c) also reveal that the impurity-induced
resonant states have wave functions extending very far along the nodal
directions of the $d$-wave OP. This extension can lead to couplings
between the impurity-induced states localized in the neighboring
super-cells. But such couplings can only broaden the resonant peaks
into narrow bands, and can not cause repulsion between the two
impurity-induced peaks.

In summary, we have studied the subtle effects of adding a very low
concentration of unitary non-magnetic impurities on the LTDOS in the
vicinity of an impurity, if the system is in the FFLO state in a d-SC.
A impurity in such a situation is locally stable at either an OP
saddle point or an OP extrema.  If the impurity is at an OP saddle
point, then the LTDOS is practically unaffected by the impurity, with
no impurity-induced resonant peak or peaks appearing, unlike in a
uniform d-SC.  If the impurity is at an OP extremum, then a
$\pm\epsilon_0 +$ (Zeeman energy) pair of finite-energy resonant
peaks are induced in the LTDOS by the impurity, instead of one at
near zero energy, as in a uniform d-SC. The physics underlying these
results has been expounded.  Confirming these results can provide
one of the strongest evidences for the existence of the FFLO state
in a d-SC.

This work is supported by a grant from the Robert A. Welch Foundation
under NO. E-1146 and by the Texas Center for Superconductivity at the
University of Houston through the State of Texas.


\begin{thebibliography}{20}
\bibitem{note1} A third type can potentially appear at surfaces and
interfaces which is not considered in this study.
\bibitem{note2} The percentage difference in excess energy due to the
impurity is about 1\% for $U_0=100$, and about 6\% for $U_0=5$.
\end{thebibliography}
\end{document}